\documentclass[12pt]{article}            
\usepackage{amssymb}
\newlength{\dinwidth}
\newlength{\dinmargin}
\begin{document}
\title{Global Properties of Vacuum States\\ in de Sitter Space}
\author{ {\sc H.J.\ Borchers} \ and \ 
 {\sc D.\ Buchholz}
\\[14pt]
{\normalsize  Institut f\"ur Theoretische Physik, Universit\"at G\"ottingen}\\
{\normalsize  D-37073 G\"ottingen, Germany}}
\date{        }
\maketitle
\renewcommand{\theequation}{\thesection.\arabic{equation}}
\newtheorem{Definition}{Definition}[section]
\newtheorem{Theorem}[Definition]{Theorem}
\newtheorem{Proposition}[Definition]{Proposition}
\newtheorem{Lemma}[Definition]{Lemma}
\newtheorem{Corollary}[Definition]{Corollary}
\newcommand{\nin}{\noindent}
\def\CA{{\cal A}}
\def\CB{{\cal B}}
\def\CCC{{\cal C}}
\def\CH{{\cal H}}
\def\CM{{\cal M}}
\def\CN{{\cal N}}
\def\CR{{\cal R}}
\def\CS{{\cal S}}
\def\CO{{\cal O}}
\def\CU{{\cal U}}
\def\CW{{\cal W}}
\def\CZ{{\cal Z}}
\def\AO{{\CA ( \CO )}}
\def\AW{{\CA ( \CW )}}
\def\CC{{\Bbb C}}
\def\RR{{\Bbb R}}
\def\I{{\rm i}}
\def\E{{\rm e}}
\def\dS{{\CS^n}}
\def\dSG{{\mbox{\it SO\/}_0(1,n)}}
\def\BW{{\Lambda_{\cal W}}}
\def\sabsatz{\par\smallskip\noindent}
\def\mabsatz{\par\medskip\noindent}
\def\babsatz{\par\vskip 1cm\noindent}
\def\tabsatz{\topinsert\vskip 1.5cm\endinsert\noindent}
\def\newline{\hfil\break\noindent}
\def\newpage{\vfill\eject\noindent}
\def\Imt{\Im m\,}
\def\ad{{\rm Ad}\,}
\def\Box{{\vcenter{\vbox{\hrule width4.4pt height.4pt 
              \hbox{\vrule width.4pt height4pt\kern4pt\vrule width.4pt}
               \hrule width4.4pt }}}}
\def\frac#1#2{{#1\over#2}}
\def\B1{{\mathchoice {\rm 1\mskip-4mu l} {\rm 1\mskip-4mu l}
  {\rm 1\mskip-4.5mu l} {\rm 1\mskip-5mu l}}}
\def\BR{{\rm I\!R}} 
\def\ch{{\rm ch}\,}
\def\sh{{\rm sh}\,}

\noindent
{\small  {\bf Abstract:} Starting from the assumption that vacuum states in
de Sitter space look for any geodesic observer like 
equilibrium states with some {\em a priori} arbitrary temperature, 
an analysis of their global properties 
is carried out in the algebraic framework of local quantum physics. 
It is shown that these states have the Reeh--Schlieder 
property and that any primary vacuum state is also pure and weakly
mixing. 
Moreover, the geodesic temperature of vacuum states has to be 
equal to the Gibbons--Hawking temperature and this 
fact is closely related to the existence of a discrete PCT--like 
symmetry. It is 
also shown that the global algebras of observables in vacuum
sectors have the same structure as their counterparts in Minkowski 
space theories.}
${}$ \\[14pt]
\section{Introduction}
\setcounter{equation}{0}
It is a well established fact that the most elementary states in 
de Sitter space, corresponding to vacuum states in Minkowski 
space, look for any geodesic observer like thermal states with 
a certain specific temperature which depends on the radius of the space. 
This universal (model independent) feature can be traced back to 
the Unruh effect \cite{Un}, the thermalizing effects of event horizons
\cite{GiHa,BiWi,Se,NaPeTh} or to stability  properties of the 
elementary states which manifest themselves in the form of 
specific analyticity properties \cite{BrMo,BrEpMo}. 

In the present article we take this characteristic feature of elementary
states (called vacuum states in the following) as input in a
general analysis of their global properties. These properties
were recently
also discussed in \cite{BrEpMo}. In the
present analysis, which is carried out in the algebraic framework of local
quantum physics \cite{Ha}, we reproduce the results in \cite{BrEpMo} 
under slightly less restrictive assumptions and exhibit further
interesting properties of vacuum states in de Sitter space which 
closely resemble those of their counterparts in Minkowski
space. 

Following is a brief outline of our results: In Sec.\ 2 we collect   
some basic properties of  de Sitter space, the de Sitter group and of 
the unitary representations of this group. After these preparations we
state in Sec.\ 3 our assumptions and establish a Reeh--Schlieder
theorem for vacuum states. In Sec.\ 4 we show that the global algebras of
observables are, in any vacuum sector, of type I according to
the classification of Murray and von Neumann and have an abelian
commutant. In particular, any primary vacuum state is also pure 
and weakly mixing. Invariant means of 
local observables with respect to certain specific one--parameter subgroups of 
the de Sitter group and their relation to the center of the 
global algebras are discussed in Sec.~5. Finally, we establish in 
Sec.\ 6 a PCT theorem in de Sitter space and present an argument 
showing that the temperature of de Sitter space
has to be equal to the Gibbons--Hawking temperature.
The article concludes with a brief summary. 

\section{De Sitter space and de Sitter group}
\setcounter{equation}{0}

For the convenience of the reader we compile here some relevant   
properties of the de Sitter space and the de Sitter group 
as well as some information on the continuous 
unitary representations of this group.  
(For an extensive
list of references on this subject cf.\ \cite{Mi}.)

The $n$--dimensional de Sitter space $\dS$ can conveniently be
described in the $n+1$--dimensional ambient Minkowski space
$\RR^{n+1}$. Assuming that $n > 1$,  
it corresponds to a one--sheeted 
hyperboloid which, in proper coordinates,
is given by 
\begin{equation}
\dS = \{ x \in \RR^{n+1} : \, x_0^2 - x_1^2 \, \dots - x_n^2 = -1 \}. 
\end{equation}
The metric and causal structure on $\dS$ are induced by the Minkowskian metric
on $\RR^{n+1}$. Accordingly, the isometry group of $\dS$ is the 
group $O(1,n)$, called the de Sitter group. Its action on $\dS$ is
given by the familiar action of the Lorentz group in the
ambient space. We restrict attention here to the
identity component of $O(1,n)$ which is usually denoted by  
$\dSG$. 

In the following we deal with certain distinguished subregions $\CW \subset
\dS$, called
wedges. These wedges are defined as the causal completions of timelike
geodesics in $\dS$. Thus they are those parts of de Sitter space which
are both, visible and accessible 
for observers moving along the respective geodesics. 
Each wedge $\CW$ can be represented as intersection of de
Sitter space $\dS$ with a wedge shaped region in the ambient space,
such as 
\begin{equation}
\CW_j = \{  x \in \RR^{n+1} : x_j > | x_0 | \, \} \, \cap \, \dS, \ \
j=1, \dots n. 
\end{equation}
We note that any wedge
$\CW \subset \dS$ is obtained from a fixed one, 
say $\CW_1$,  by the action of some 
element of $\dSG$. Moreover, the spacelike
complement $\CW \, '$ of a wedge $\CW$ is again a wedge. 

Given a wedge $\CW$ there is a unique  one--parameter subgroup of 
$\dSG$ which leaves $\CW$ invariant and 
induces a future directed Killing vector field in that
region. We denote this group by $\BW (t), \, t \in \RR$, and call it
the group of boosts associated with $\CW$. It describes the
time evolution for the geodesic observer in $\CW$. 
The causal complement $\CW \, '$ of $\CW$ is also invariant under the 
action of $\BW (t), \, t \in \RR$, but the corresponding Killing 
vector field is past directed in that region. Hence there holds
$\Lambda_{\CW \, '} (t) = \BW (-t), \, t \in \RR$.

Let us now turn to a discussion of the continuous unitary
representations of $\dSG$. Given any such representation $U$ on some 
Hilbert space $\CH$ we denote the corresponding selfadjoint generators
with respect to the chosen coordinate system  
by $M_{\mu \nu}, \ \mu,\nu = 0, 1, \dots n$. They satisfy on a
canonical 
domain of analytic vectors \cite{Ne} the Lie--algebra relations
\begin{equation}
[M_{\mu \nu}, M_{\rho \sigma}]=-\,i g_{\mu \rho} M_{\nu \sigma}
+\, i g_{\mu \sigma} M_{\nu \rho}+\, i g_{\nu \rho}  M_{\mu \sigma}
-\, i g_{\nu \sigma} M_{\mu \rho}, 
\end{equation}
where $g_{\mu \nu}$ is the metric tensor of the ambient Minkowski
space. The operators
$M_{0 j}$ generate the action of the boosts
associated with the wedges $\CW_j$, 
\begin{equation}
U( \Lambda_{\CW_j} (t)) = \E^{\, it M_{0 j}}, \ \ t \in \RR, \ j=1, \dots n,
\end{equation}
and the $M_{j k}, \, j,k=1, \dots n$, are the generators of spatial
rotations. 

For fixed
$j\neq k$ the operators $M_{0 j}, \, M_{0 k}$ and $M_{j k}$ form a Lie 
sub--algebra and there holds for $s,t\in\RR$
\begin{eqnarray} 
\E^{\, is M_{0 j}} \, \E^{\, it M_{0 k}} \, \E^{-\, is M_{0 j}} & = & 
\E^{\, it(\ch(s) M_{0 k}+\sh(s) M_{j k})} \label{2.5} \\
\E^{\, is M_{j k}} \, \E^{\, it M_{0 k}} \, \E^{-\, is M_{j k}} & = &
\E^{\, it (\cos(s) M_{0 k}+\sin(s) M_{0 j})}. \label{2.6}
\end{eqnarray}  
These relations are repeatedly used in the proofs of the following results.

\begin{Lemma}
Let $\CN\subset \dSG$ be any open neighborhood of the unit
element in $\dSG$ and let $\BW (t), \, t\in\RR$, be the 
boosts associated with a given wedge $\CW$.
Then the strong closure of the group generated by the unitary operators   
$U(\Lambda \BW (t) \Lambda^{-1})$ with $t\in\RR$, 
$\Lambda\in\CN$, coincides with $U(\dSG)$. 
\end{Lemma}
{\em Proof\/}: Because of the de Sitter invariance of the problem we can  
assume without loss of generality that $\CW $ is the wedge
$\CW_1$. The statement can then be established by the 
following computation. 
Let $U_\CN$ be the closed unitary group generated by 
$U(\Lambda \Lambda_{\CW_1} (t) \Lambda^{-1})$ with $t\in\RR$ and
$\Lambda\in\CN$. 
It follows from (\ref{2.5}) that for sufficiently
small $|s|$ and any $t\in\RR$ there holds
$\E^{it(\ch(s)M_{0 1}+\sh(s)M_{j 1})}\in U_\CN$ for $j=2, \dots n$.
Keeping $s\neq0$ fixed one sees by an application of the Trotter
product formula \cite{ReSi} to the product of the one--parameter groups
$\E^{it(\ch(s)M_{0 1}+\sh(s)M_{j 1})}$ and $\E^{-it\ch(s)M_{0 1}}$
that the rotations $\E^{it\sh(s)M_{j 1}},\, t\in\RR, \, j=2, \dots n$, belong
to $U_\CN$. Relation (\ref{2.6}) then implies that also 
$\E^{itM_{0 j}} \in U_\CN$ for $t\in\RR, \, j=1, \dots n$. Since these 
operators generate $U(\dSG)$ there holds $U_\CN=U(\dSG)$,
proving the statement.$\hfill\blacksquare$
 
\begin{Lemma} Let $\Psi\in\CH$ be invariant
under the action of $U(\BW (t)),\, t\in\RR$, where $\BW (t), \,
t\in\RR$, 
is the group of boosts associated with a given wedge $\CW$.
Then $\Psi$ is invariant under the
action of $U(\dSG)$. \label{2.2}
\end{Lemma}
{\em Proof\/}: As in the proof of the preceding lemma we may
assume without restriction of generality that $\CW $ is the wedge
$\CW_1$. Putting $t=2r\E^{-|s|}$ in relation (\ref{2.5}) it follows from 
the continuity of the representation $U$ 
that in the sense of strong operator convergence on $\CH$
\begin{equation}
\lim\limits_{s\to\pm\infty}
\E^{isM_{0 1}} \E^{2 ir\E^{-|s|}M_{0 j}} \E^{-isM_{0 1}}=
\E^{ir(M_{0 j}\pm M_{1 j})}
\end{equation}
for $j=2, \dots n$. 
On the other hand, since $\Psi$ is invariant under 
the action of the unitary operators $\E^{isM_{0 1}}, \ s \in \RR$, and since
$\E^{2ir\E^{-|s|}M_{0 j}}$ converges to $1$ in the strong operator 
topology for $s \rightarrow \pm \infty$ and fixed $r$, we get 
\begin{equation}
\lim_{s \rightarrow \pm \infty} || \E^{ is M_{0 1}} 
\E^{2ir\E^{-|s|}M_{0 j}} \E^{-is M_{0 1}} \Psi - \Psi || =
\lim_{s \rightarrow \pm \infty} || \E^{2ir\E^{-|s|}M_{0 j}}  \Psi -
\Psi || = 0.
\end{equation}
Combining these relations we obtain  
\begin{equation}
\E^{ir(M_{0 j} \pm M_{1 j})} \Psi = \Psi \ \ \mbox{for} \ \ r \in  \RR, \,
j = 2, \dots n. 
\end{equation}
By a similar argument as in the proof of the preceding lemma it then follows 
that $U(\Lambda) \Psi = \Psi$ for any $\Lambda \in \dSG$.$\hfill\blacksquare$

We conclude this section by recalling a result of Nelson \cite{Ne} on
the existence of analytic vectors for generators 
of unitary representations of Lie--groups. We state this result in a
form which is convenient for the subsequent applications. 

\begin{Lemma} Let $\CCC$ be a sufficiently small 
neighborhood of the origin in 
$\CC$. There exists a dense set of vectors $\Phi \in \CH$ 
such that 
\begin{equation} 
 \sum_{n=0}^\infty \, 
\frac{||(u M_{0 k} + v M_{j k})^n \, \Phi ||}{n!} <
\infty 
\end{equation}
for $u,v \in \CCC$ and $j,k = 1, \dots n$. Phrased
differently, the vectors $\Phi$ are analytic for the respective 
generators with a uniform radius of convergence. \label{2.3}
\end{Lemma}
{\em Proof\/}: The statement follows 
from Theorem 3 in  \cite{Ne} by taking
also into account the quantitative estimates in Corollary 3.1 and
Lemma 6.2 of that reference.$\hfill\blacksquare$
 
\section{Reeh--Schlieder property of vacuum states}
\setcounter{equation}{0}

Before we turn now to the analysis of vacuum states in de Sitter
space we briefly list our assumptions, establish our notation and add
a few comments.  

1.\ {\em (Locality)\/} There is an inclusion preserving mapping  
\begin{equation}
\CO \rightarrow \AO
\end{equation}
from the set of open, bounded, contractible regions $\CO \subset \dS$ 
to von Neumann algebras $\AO$ on 
some Hilbert space $\CH$. We interpret each $\AO$ as the 
algebra generated by all observables which can be measured in $\CO$.
For any wedge $\CW \subset \dS$ the 
corresponding algebra $\AW$ is defined as the von Neumann 
algebra generated by the local algebras $\AO$ with $\CO \subset \CW$,    
and $\CA$ denotes the von Neumann algebra generated by all local 
algebras $\AO$. The local algebras are supposed to
satisfy the condition of locality, i.e.
\begin{equation}
\CA ( \CO_1 ) \subset \CA ( \CO_2)' \ \ \mbox{if} \ \ \CO_1 \subset \CO_2',
\end{equation} 
where $\CO'$ denotes the spacelike complement of $\CO$ in $\dS$ and 
$\AO'$ the commutant of $\AO$ in $\CB (\CH) $. 

2.\ {\em (Covariance)\/} On $\CH$ there is a continuous unitary
representation $U$ of the de Sitter group $\dSG$ which induces
automorphisms $\alpha$ of $\CB (\CH)$ acting covariantly on 
the local algebras. More concretely, putting  
$\alpha_{\Lambda} (\, \cdot \,) \doteq U(\Lambda) \, \cdot \, U(\Lambda)^{-1} 
, \, \Lambda \in \dSG$, there holds for each region $\CO \subset \dS$ 
\begin{equation}
\alpha_\Lambda (\AO) =
\CA (\Lambda \CO). 
\end{equation} 

3.\ {\em (de Sitter Vacuum)\/} There is a unit vector 
$\Omega \in \CH$, describing the vacuum,  
which is invariant under the action of $U(\dSG)$ and cyclic
for the global algebra $\CA$. The corresponding vector state $\omega$
on $\CA$, given by 
\begin{equation}
\omega (A) = (\Omega, \, A \, \Omega), \ \ A \in \CA, \label{3.4}
\end{equation} 
has the following {\em geodesic KMS--property\/} suggested by the
results of Gibbons and Hawking \cite{GiHa}: 
For every wedge $\CW$ the restriction (partial state)  
$\omega \! \upharpoonright \! \AW$ 
satisfies the KMS--condition at some inverse temperature $\beta > 0$
with respect to the time evolution (boosts) $\BW (t), \, t \in \RR$, 
associated with $\CW$. In other words, for any pair of operators 
$A,B \in \AW$ there exists an analytic
function $F$ in the strip 
$\{ z \in \CC : 0 < \mbox{Im} z < \beta \}$ with continuous
boundary values at $\mbox{Im} z = 0$ and $\mbox{Im} z = \beta$, which
are given respectively (for $t \in \RR$) by 
\begin{equation}
F(t) = \omega (A \alpha_{\BW (t)} (B) ), \ \ 
F(t + i\beta) = \omega (\alpha_{\BW (t)} (B) A). 
\end{equation}
In order to cover also the case of degenerate vacuum states we do
{\em not} assume here that the vacuum vector $\Omega$ is (up to a
phase) unique. 

The inverse temperature $\beta$ in the 
preceding condition has to be
the same for all wedges $\CW$ because of the invariance of $\Omega$ 
under the action of the de Sitter group. Its actual value  
has been determined by several authors in a general setting by starting
from various assumptions, such as the
condition of local stability \cite{HaNaSt,NaPeTh}, the weak spectral
condition \cite{BrEpMo} or the condition of modular covariance on
lightlike hyper-surfaces \cite{VeSu}. 
As we shall see, the present assumptions
already fix the value of $\beta$. 

Our last condition expresses the idea that all observables are 
built from strictly local ones. It is a  
standard assumption in the case of Minkowski space theories. 

4. {\em (Weak additivity)\/} For each open region $\CO \subset \dS$ 
there holds 
\begin{equation}
\bigvee_{\Lambda \in \mbox{\scriptsize \it SO\/}_0(1,n)} 
\CA ( \Lambda \CO)  = \CA.
\end{equation}

Note that, for $n > 1$,  
$\{ \Lambda \CO : \Lambda \in \dSG \}$ defines a covering 
of $\dS$ since $\dSG$ acts transitively on that space. So the 
condition is clearly satisfied if the local algebras 
are generated by Wightman fields. 

We turn now to the analysis of the 
cyclicity properties of $\Omega$ with respect to the local 
algebras $\AO$. 

\begin{Definition}
Let $\CO \subset \dS$ be any open  region. The $*$--algebra
$\CB(\CO)$ is defined as the set of operators  $B \in \AO$ 
for which there exists some  
neighborhood $\CN \subset \dSG$ of the unit element in $\dSG$ 
(depending on $B$) such that 
\begin{equation}
\alpha_{\Lambda}(B)\in\AO \quad\mbox{\it for}\quad \Lambda \in \CN.
\end{equation}
\end{Definition}
It is apparent that $\CB(\CO)$ is indeed a $*$--algebra and that $\CA(\CO_0)
\subset \CB(\CO)$ for any region $\CO_0$ 
whose closure satisfies $\overline{\CO_0}\subset \CO$.

In the subsequent lemmas we establish some technical properties of the
orthogonal complements of the spaces $\CB(\CO) \Omega$ in $\CH$.
(Cf.\ \cite{DrSuWi} for a similar discussion in case of 
Minkowski space theories.)  
It suffices for our purposes to consider regions $\CO \subset \dS$ 
which are so small that there exists a wedge $\CW$ 
and an open neighborhood $\CN\subset \dSG$ of the unit
element in $\dSG$ such that $\Lambda^{-1} \CO \subset \CW$ for all 
$\Lambda \in \CN$. 
Then, if  $\BW (t),\, t\in\RR$, is the one--parameter group of boosts 
associated with $\CW$, there holds 
\begin{equation}
\Lambda \BW(t) \Lambda^{-1} \CO \subset \Lambda \CW 
\ \ \mbox{for} \ \ \Lambda\in\CN, \ t\in\RR,  \label{3.8}
\end{equation}
where $\Lambda \BW(t) \Lambda^{-1}, \ t \in \RR$, are the boosts
associated with the wedge $\Lambda \CW$. 
 
\begin{Lemma}
Let $\CO \subset \dS$ be a sufficiently small region 
(in the sense described above) and 
let $\Psi \in \CH$ be a vector with the property that
\begin{equation}
(\Psi, B \, \Omega)=0\quad \mbox{\it for}\quad B\in \CB(\CO). 
\end{equation} 
Then the vectors $U(\Lambda)\Psi$, $\Lambda \in \dSG$, have the same property.
\end{Lemma}
{\em Proof\/}: 
Let $B\in\CB(\CO)$ and $\Lambda\in\CN$ with $\CN$ as in relation (\ref{3.8}). 
It follows from the definition of
$\CB(\CO)$ and the continuity of the boosts that there is an $\varepsilon > 0$
such that 
$\alpha_{\Lambda \BW (t) \Lambda^{-1}}(B)\in \CB(\CO)$
for $|t|<\varepsilon$ and consequently  
\begin{equation}
(\Psi,\alpha_{\Lambda \BW (t) \Lambda^{-1}}(B) \, \Omega)=0 
\ \ \mbox{for} \ \ |t|<\varepsilon. \label{3.10}
\end{equation}
On the other hand, since 
$\Lambda \BW(t) \Lambda^{-1} \CO \subset \Lambda \CW$, 
there holds
$\alpha_{\Lambda \BW (t) \Lambda^{-1}}(B)\in \CA(\Lambda \CW) $ 
for $t\in\RR$. So the geodesic KMS--property 
of $\Omega$ implies \cite{KaRi} that
\begin{equation}
t\longrightarrow \alpha_{\Lambda \BW (t) \Lambda^{-1}}(B) \,
\Omega,\quad t\in\RR
\end{equation}
extends analytically to some vector--valued function in the strip
$\{ z \in \CC : 0 < \mbox{Im} z < \beta/2 \}$. Combining these two 
informations it follows that 
\begin{equation}
(U(\Lambda \BW (t) \Lambda^{-1})\Psi, B \, \Omega) = 
(\Psi, \alpha_{\Lambda \BW (-t) \Lambda^{-1}}(B) \, \Omega)=0\quad
\end{equation}
for all $t\in\RR$ and $B \in \CB(\CO)$. 
Since $\Lambda\in\CN$ was arbitrary, we conclude by repetition of the 
preceding argument that for any $\Lambda_1, \dots \Lambda_k\in\CN$ and
$t_1, \dots t_k\in\RR$
\begin{equation}
(U(\Lambda_1 \BW (t_1) \Lambda_1^{-1}) \dots 
U(\Lambda_k \BW (t_k) \Lambda_k^{-1})  \Psi, B \,  \Omega)=0.
\end{equation}
As $U(\dSG)$ is generated by products of the boost 
operators $U(\Lambda \BW (t) \Lambda^{-1})$ with  
$\Lambda\in\CN, \, t\in\RR$, cf.\ Lemma 2.1, the assertion follows. 
$\hfill\blacksquare$

\begin{Lemma} 
Let $\CO \subset \dS$ and $\Psi \in \CH$ be as in the preceding
lemma. There holds
for $\Lambda_1, \dots \Lambda_k\in\dSG$ and $B_1, \dots B_k\in \CB(\CO)$
\begin{equation}
(\Psi, \alpha_{\Lambda_1}(B_1) \cdots \alpha_{\Lambda_k}(B_k) \, \Omega)=0.
\end{equation}
\end{Lemma}
{\em Proof\/}: As $\CB(\CO)$ is a $*$--algebra we see from the preceding 
lemma that $B^* U(\Lambda) \, \Psi$ is orthogonal to $\CB(\CO) \, \Omega$
for any $B \in \CB(\CO)$ and $\Lambda \in \dSG$. 
So the statement follows by induction.$\hfill\blacksquare$ 

We are now in a position to establish the Reeh--Schlieder property of 
$\Omega$, i.e. the fact that $\Omega$ is a cyclic vector for all local
algebras. 

\begin{Theorem}
For any open region $\CO \subset \dS$ there holds 
\begin{equation}
\overline{\AO \, \Omega} = \CH.
\end{equation}
\end{Theorem}
{\em Proof\/}: 
We may assume that $\CO$ is so small that the preceding
lemma can be applied. Now if $\Psi \in \CH$ is orthogonal to
$\AO \, \Omega$ it is also orthogonal to $\CB(\CO) \, \Omega$ and
consequently to 
$
\left( \bigvee_{\Lambda \in \mbox{\scriptsize \it SO\/}_0(1,n)} 
\alpha_\Lambda (\CB(\CO)) \right) \Omega.
$
As $\CA (\CO_0) \subset \CB(\CO)$ if 
$\overline{\CO_0}\subset \CO$ there holds 
\begin{equation}
\bigvee_{\Lambda \in \mbox{\scriptsize \it SO\/}_0(1,n)} 
\alpha_\Lambda (\CB(\CO)) \supset  
\bigvee_{\Lambda \in \mbox{\scriptsize \it SO\/}_0(1,n)} 
\alpha_\Lambda (\CA (\CO_0))  = 
\bigvee_{\Lambda \in \mbox{\scriptsize \it SO\/}_0(1,n)} 
\CA (\Lambda \CO_0) = \CA, 
\end{equation}
where in the last equality we made use 
of weak additivity. Since $\Omega$ is cyclic for $\CA$ it follows that
$\Psi = 0$, completing the proof. $\hfill\blacksquare$

\section{Type of the global algebra $\CA$}
\setcounter{equation}{0}

We turn now to the analysis 
of the global algebra $\CA$, where we will 
make use of modular theory, cf.\ for example \cite{KaRi}. 
The geodesic KMS--property implies that 
$\alpha_{\BW (-t)}, \, t \in \RR$, is (apart from a rescaling of 
the parameter $t$ by $\beta$) the group of modular automorphisms
associated with the pair $\{ \AW, \Omega \}$ for any wedge $\CW
\subset \dS$. Hence, by the basic results of 
modular theory, $\alpha_{\BW (t)}, \, t \in \RR$, is the 
modular group of $\{ \AW ', \, \Omega \}$. This 
fact will be used at various points in the subsequent investigation.

We begin our discussion with two preparatory  
propositions which are of interest in 
their own right.

\begin{Proposition} The commutant $\CA '$ of $\CA$ is pointwise
invariant under the adjoint action of $U(\Lambda)$, $\Lambda \in \dSG$,
i.e.\ the representation $U$ of the de Sitter group is contained 
in the global algebra of observables $\CA$. \label{4.1}
\end{Proposition}
{\em Proof\/}: We fix a wedge $\CW $ and consider the corresponding
automorphisms 
$\alpha_{\BW (t)}, \, t \in \RR$. As $\CW$ is 
invariant under the boosts $\BW (t)$  the algebras
$\AW '$ and $\AW ' \cap \CA $ are invariant under the action of 
these automorphisms. Both algebras contain $\CA (\CW ')$ and thus
have $\Omega$ as a cyclic vector. 

If $X \in \CA' \subset \AW '$ and $A \in \AW'  \cap \CA$ 
it follows from modular theory that the function
$t \rightarrow (\Omega, A \alpha_{\BW (t)}(X) \, \Omega) $
extends to a bounded analytic function $F$ in the strip 
$\{ z \in \CC : 0 > \mbox{Im} z > - \beta \}$. Moreover, the boundary value 
of $F$ at $\mbox{Im} z = - \beta$ is given by 
$F(t-i\beta) = (\Omega, \alpha_{\BW (t)}(X) A \, \Omega) = 
(\Omega, A \alpha_{\BW (t)}(X) \, \Omega)$, where in the second
equality we have used the commutativity of $\CA '$ and $\AW ' \cap
\CA$. Hence $F$ can be extended by periodicity to a bounded 
analytic function on $\CC$ and thus is constant. Since 
$A \in \AW ' \cap \CA$ is arbitrary and $\Omega$ is cyclic for 
$\AW ' \cap \CA$ and separating for $\CA '$ we conclude that 
$X = \alpha_{\BW (t)}(X) $, i.e.\ $X$ commutes with the unitaries
$U(\BW(t))$ for $t \in \RR$ and every wedge $\CW$. 
As these unitaries generate the group $U(\dSG)$, 
the proof is complete.$\hfill\blacksquare$ 
 
\begin{Proposition}
Let $\CO \subset \dS$ be any open region and let $E_0$ be the projection
onto the space of $U(\dSG)$--invariant vectors in $\CH$. Then the 
von Neumann algebra generated
by $E_0$ and $\AO$ coincides with $\CA$. \label{4.2}
\end{Proposition}
{\em Proof\/}: 
Given $\CO \subset \dS$ we pick another open 
region $\CO_0$ such that for 
some neigbourhood $\CN$ of the unit element of $\dSG$ there holds
$\Lambda \CO_0 \subset \CO$ for $\Lambda\in\CN$.
Now let $C \in \{ \AO, \, E_0 \}'$. Then there holds for any 
$A\in\CA(\CO_0)$ and $\Lambda\in\CN$
\begin{equation}
U(\Lambda)^{-1} C U(\Lambda) A \, \Omega = 
A U(\Lambda)^{-1} C E_0 \, \Omega = 
A U(\Lambda)^{-1} E_0 C \, \Omega = 
A C \,\Omega =
C A \, \Omega.
\end{equation}
Since $\Omega$ is cyclic for $\CA (\CO_0)$ it follows that 
$U(\Lambda)^{-1} C U(\Lambda) = C $
for $\Lambda\in\CN$ and therefore for  
$\Lambda\in\dSG$. Hence $C$ commutes also 
with $\bigvee_{\Lambda \in \mbox{\scriptsize \it SO\/}_0(1,n)} 
\CA (\Lambda \CO_0) =  \CA $,
where we have used weak additivity, 
and consequently $ \{E_0, \CA (\CO) \}' \subset \{E_0, \CA \}'$. But  
$E_0 \in U(\dSG){''} \subset \CA$,  where the inclusion follows
from the  preceding proposition. Hence 
$\CA \subset \{E_0, \CA (\CO) \}{''} \subset \CA$ as claimed.
$\hfill\blacksquare$

With this information we can now establish the following theorem.

\begin{Theorem} In the vacuum sector of any de Sitter theory 
there holds:\newline
{$(a)$} The commutant $\CA'$ of $\CA$ is abelian (i.e.\ $\CA$
is of type I and $\CA'$ is the center of $\CA$).\newline
{$(b)$} The projection $E_0$ onto the space of all
$U(\dSG)$--invariant vectors in $\CH$ is an abelian projection in $\CA$
with central support $1$. \label{4.3}
\end{Theorem}
{\em Proof\/}: Let $\CW$ be any wedge and let $X_1,X_2\in \CA \, '
\subset\AW '$,  $A\in\AW '\cap\CA$. 
As in the proof or Proposition \ref{4.1} 
we make use of the modular theory for $\{\AW ', \Omega \}$ and
consider the function 
$t \rightarrow (\Omega,AX_1 \alpha_{\BW(t)}(X_2) \, \Omega)$. It
extends to an analytic function $F$ in the strip 
$\{ z \in \CC : 0 > \mbox{Im} z > - \beta \}$ whose boundary value 
at $\mbox{Im} z = - \beta$ is given by 
$F(t-i\beta) = (\Omega, \alpha_{\BW(t)}(X_2) A X_1 \, \Omega)$. 
On the other hand, the pointwise invariance of $\CA '$ 
under the action of $\alpha_{\BW(t)}$, cf.\ Proposition \ref{4.1}, implies
that $F$ is constant. Combining these two facts we get    
\begin{equation}
(\Omega,AX_1X_2 \, \Omega) = (\Omega,X_2AX_1 \, \Omega) = 
(\Omega,AX_2X_1 \, \Omega), 
\end{equation}
where in the second equality we made use of the commutativity of 
$A$ and $X_2$. 
The cyclicity of $\Omega$ for $\AW '\cap\CA$ implies
$[X_1,X_2] \, \Omega=0$. Since $\Omega$ is separating for $\CA '$ it 
follows that $[X_1,X_2] \, =0$. So $\CA '$ is abelian, proving 
the first part of the statement.

For the proof of the second part we 
pick a wedge $\CW$ and choose $A\in\AW$ and $B\in \CA (\CW \, ')$. 
By the mean ergodic theorem (see e.g.\ \cite{Ka}) there holds 
in the sense of strong operator convergence
\begin{equation} 
\lim\limits_{T\to\infty} 
{T^{-1}} \int\limits_0^T dt 
\, U(\BW( \pm t)) = F_0,
\end{equation}
where $F_0$ denotes the projection onto the subspace of vectors in
$\CH$ which are invariant under the action of the unitaries
$U(\BW(t))$, $t \in \RR$. Hence, making use of locality and the 
invariance of $E_0$ under left and right multiplication with  
$U(\BW(t))$, we get  
\begin{eqnarray}
E_0BF_0AE_0 = 
 \lim\limits_{T\to\infty} 
{T^{-1}} \! \int\limits_0^T \! dt 
\, E_0 B \alpha_{\BW (t)}(A)E_0 =  \nonumber \\
= \lim\limits_{T\to\infty}
{T^{-1}} \! \int\limits_0^T \! dt 
\, E_0 \alpha_{\BW (t)}(A) B E_0 = 
E_0AF_0BE_0.
\end{eqnarray}
According to Lemma \ref{2.2} $F_0$ coincides with $E_0$  and hence
\begin{equation}
E_0BE_0AE_0=E_0AE_0BE_0. \label{eq4.6}
\end{equation}
This shows that the algebras 
$\{ E_0 \AW E_0 \! \upharpoonright \! E_0 \CH \} ''$ and 
$\{ E_0 \CA (\CW \, ') E_0 \! \upharpoonright \! E_0 \CH \} ''$ commute. 
By Proposition \ref{4.2} both algebras coincide with 
$E_0 \CA E_0$, hence relation (\ref{eq4.6}) holds for all $A,B \in \CA$,
proving that $E_0$ is an abelian projection. 

Finally, let $E$ be any projection in the center of $\CA$ 
which dominates $E_0$, i.e.\ $E E_0 = E_0$. Then there holds in particular 
$E \, \Omega = \Omega$ and since $\Omega$ is separating for 
the center it follows that $E = 1$. So $E_0$ has central support
$1$.$\hfill\blacksquare$

\begin{Corollary}
The following statements are equivalent for any vacuum state $\omega$:\newline
(a) $\omega$ is a primary state\newline
(b) $\omega$ is a pure state\newline
(c) $\omega$ is weakly mixing 
with respect to the action of boosts.
\end{Corollary}
{\em Proof:\/} If $\omega$ is primary $\CA$ has a trivial center. 
But according to part (a) of 
the preceding theorem the center of $\CA$ is equal to $\CA \, '$
and consequently $\CA \, ' = \CC \, 1$. Hence $\omega$ is a pure state. 

In the latter case there holds $\CA = \CB (\CH)$ which implies  
$E_0 \CA E_0 = \CB ( E_0 \CH)$. According to part (b) of the preceding theorem 
the algebra $E_0 \CA E_0$ is abelian, so $E_0$ must be a
one-dimensional projection. Thus by the mean ergodic theorem
and Lemma 2.2 
\begin{equation}
\lim\limits_{T\to\infty} 
{T^{-1}}\! \int\limits_0^T \! dt 
\, \omega( B \alpha_{\BW (t)}(A) ) = (\Omega, B E_0 A \, \Omega) =
\omega(B) \omega(A) \label{4.6}
\end{equation}
for any $A,B \in \CA$ which shows that $\omega$ is weakly mixing. 
 
Conversely, relation (\ref{4.6}) implies that the projection 
$E_0 \in \CA$ is one--dimensional. Hence if  $X \in \CA \, '$
there holds $X \, \Omega = E_0 X \, \Omega = \omega(X) \,
\Omega$. Since $\Omega$ is separating for $\CA \, '$ it follows that 
$\CA \, ' = \CC \, 1$. So the state $\omega$ is pure and {\em a fortiori\/}
primary.$\hfill\blacksquare$ 

\section{Invariant means and the center of $\CA$}
\setcounter{equation}{0}

We analyze now   
the properties of the invariant means 
on $\CB (\CH)$ which are induced by the 
adjoint action of the boost operators $U(\BW (t)), \, t \in \RR$, 
associated with arbitrary wedges $\CW \subset \dS$. 
Since $\RR$ is amenable such means
exist in the space of linear 
mappings on $\CB (\CH)$ as limit points of the nets 
\begin{equation} 
{T^{-1}} \! \int\limits_0^T \! dt \,  
U(\BW (t)) \, \cdot \, U(\BW (t))^{-1}, \quad T \rightarrow \infty, 
\end{equation}
in the so--called point--weak--open topology. We denote the respective limits 
by $M_\CW$
and note that they are, for given $\CW$, in general neither unique 
nor normal. Therefore the following result is of some interest. 

\begin{Proposition} Let $\CW$ be any wedge and let $M_\CW$ be a 
corresponding mean on $\CB(\CH)$. The  
restriction $M_\CW \! \! \upharpoonright \! \!  \AW$
is unique, normal, and its range lies in $\AW$ and 
coincides with the center of $\CA$.
\end{Proposition}
{\em Proof\/}: Let $A \in \AW$. 
{}From the invariance of $\AW$ under the adjoint action of 
$U(\BW (t)), \, t \in \RR$, it follows 
that $M_\CW (A)$ belongs to $\AW$
and commutes with the unitary operators  $U(\BW (t)), \, t \in \RR$. So 
Lemma \ref{2.2} implies that $M_\CW (A) \, \Omega \in E_0 \CH$.   
Now let $\CN \subset \dSG$ be 
a neighbourhood of the unit element of the de Sitter group 
such that $\Lambda \CW \cup \CW$ has
an open spacelike complement for 
$\Lambda \in \CN$. Then, 
because of locality and the Reeh--Schlieder property, 
$\Omega$ is separating for $\CA(\Lambda \CW) \bigvee \AW$. 
Moreover, by the mean ergodic theorem and Lemma \ref{2.2},  
$U(\Lambda) M_\CW (A) U(\Lambda)^{-1} \, \Omega = E_0 A \, \Omega
= M_\CW (A) \, \Omega$. So there holds 
$\alpha_{\Lambda} (M_\CW (A)) = M_\CW (A)$
for $\Lambda \in \CN$ and consequently for all $\Lambda\in \dSG$. 
As the operators $M_\CW (A), \, A \in \AW$, commute with $\CA (\CW \, ')$ it 
follows that they also commute with 
$\bigvee_{\Lambda \in \mbox{\scriptsize \it SO\/}_0(1,n)} 
\CA ( \Lambda \CW \, ') = \CA$ 
and thus belong
to the center of $\CA$ by Theorem \ref{4.3}. 

Because of the fact that $\Omega$ is separating for the center and 
the relation 
$M_\CW (A) \Omega = E_0A\Omega$ it is then clear that 
$M_\CW \! \upharpoonright \! \AW$ is unique and normal.

The preceding results imply  
that $M_\CW (\AW)$ is contained in $\AW$ and a subset of the
center of $\CA$.  As the elements of the 
center are pointwise invariant under the action of $M_\CW$  it is also
clear that $M_\CW (\AW)$ is a von Neumann algebra. Its 
restriction to $E_0 \CH$ has $\Omega$ as a cyclic vector by the 
Reeh--Schlieder property. So it is maximally
abelian in $E_0 \CH$ 
and therefore contains the restriction of the center of $\CA$
to that space. But the center of $\CA$ is faithfully represented on
$E_0 \CH$, so the assertion follows.$\hfill\blacksquare$

We mention as an aside that it follows from this proposition that every 
wedge algebra $\AW$ is of type III${_1}$ according to the
classification of Connes. For it implies that the centralizer of
$\omega$ in $\AW$ coincides with the center. By central decomposition 
one may therefore restrict attention to the case where the wedge
algebras are factors and the centralizers are trivial. Making also use of the 
fact that the modular groups $U(\BW(t)), \, t \in \RR$, cannot be 
cyclic because of the group structure of $\dSG$ 
(unless the representation $U$ is trivial)
the assertion then follows from the well--known results 
of Connes in \cite{Co}. 

It is neither clear what can be said about the action of $M_\CW$ on  
algebras of arbitrary (even bounded) regions, nor how these means 
depend on the choice of the wedge $\CW$. Nevertheless it is possible 
to define a universal invariant mean $M$ of the operators in 
the set--theoretic union of algebras 
$\bigcup_{\CW \subset \dS} \AW$ 
with values in the center of $\CA$.
(Note that this union is neither an algebra nor a vector space.) 
We define $M$ by setting for any wedge $\CW$
\begin{equation}
M \!  \upharpoonright \!  \AW \doteq  
M_\CW \!  \upharpoonright \! \AW. 
\end{equation}
For the proof that this definition is consistent, let 
$\CW_1,\CW_2$ be wedges and let $A\in \CA (\CW_1)\cap\CA(\CW_2)$. 
Then $M_{\CW_1}(A),M_{\CW_2}(A)$ are elements of the center of $\CA$ and 
there holds $M_{\CW_1}(A) \, \Omega= E_0A \, \Omega= 
M_{\CW_2}(A) \, \Omega$. As $\Omega$ is
separating for the center, this implies $M_{\CW_1}(A)=M_{\CW_2}(A)$,
proving the consistency. 

Since for $A\in\AW$ and $\Lambda \in \dSG$ there holds 
$\alpha_\Lambda (A) \in \CA ( \Lambda \CW)$ one obtains 
$M(\alpha_\Lambda (A)) \, \Omega=
E_0 \alpha_\Lambda (A) \, \Omega=E_0A \, \Omega=M(A) \,
\Omega.$ It follows that 
$M(\alpha_\Lambda (A)) = M(A)$ for $\Lambda \in \dSG$, $A\in \AW$
and any wedge $\CW$. So we have established the following proposition.

\begin{Proposition}
There exists a unique map
\begin{equation}
M: \bigcup_{\CW \subset \dS} \AW  \longrightarrow  \mbox{center}\,(\CA) 
\end{equation}
which is invariant under the right and left action of 
$\alpha_\Lambda, \Lambda \in \dSG$, and 
whose restriction to $\AW$ coincides with the corresponding
mean $M_\CW$, $\CW \subset \dS$.
\end{Proposition}
It is probably not meaningful to extend $M$ to operators which are
localized in regions larger than wedges.

\section{PCT and the temperature of de Sitter space}
\setcounter{equation}{0}

We finally discuss the implications of the geodesic KMS--property
of vacuum states for the modular conjugations $J_\CW$
associated with the wedge algebras $\AW$ and the vacuum vector
$\Omega$. The following proposition is an easy consequence of standard
results in modular theory.

\begin{Proposition} There holds wedge duality for any wedge $\CW
  \subset \dS$,
\begin{equation}
J_\CW \AW J_\CW = \AW \, ' = \CA (\CW \, '). \label{eq6.1}
\end{equation} \label{6.1}
\end{Proposition}
{\em Proof\/}: The first equality in the statement is a basic result 
of modular theory. For the proof of the second equality it suffices 
to note that (i) $\CA(\CW\,') \subset \AW '$ because of locality, 
(ii) $\Omega$ is cyclic for $\CA(\CW\,')$ by the Reeh--Schlieder 
property and (iii) $\CA(\CW\,')$ is stable under the action
of the modular group
$\alpha_{\BW (t)}, \, t \in \RR$, associated with
the pair $\{ \AW ', \, \Omega \}$. It then follows from a well known
result in modular theory \cite[Theorem 9.2.36]{KaRi} that 
$\CA(\CW\,') = \AW '$.$\hfill\blacksquare$

We will show next that the existence of the modular
conjugations $J_\CW$ fixes  
the inverse temperature $\beta$. Moreover, the 
specific form of the adjoint action of these conjugations
on the unitary group $U(\dSG)$ can be computed. 

For the proof we consider the wedge $\CW_1$, cf.\ (\ref{2.2}),
and the corresponding 
modular group $\E^{-it M_{01}}, \, t \in \RR$, and conjugation
$J_{\CW_1}$ associated with 
$\CA (\CW_1)$ and $\Omega$. 
We also pick a region $\CO \subset \CW_1$ such that 
$\Lambda \CO \subset \CW_1$ for all $\Lambda$ in some neighborhood 
of the unit element in $\dSG$. Thus, for 
sufficiently small $s \in \RR$, there holds 
\begin{equation}
\E^{is M_{0j}} \CA (\CO) \E^{-is M_{0j}} \subset {\CA}(\CW_1), \label{6.2}
\end{equation}
where 
$\E^{is M_{0j}}$ are the boost operators associated 
with the wedges $\CW_j, \, j = 1, \dots n$.

Making use of relation (\ref{2.5}) we get for $A \in \CA (\CO)$
and $j = 2, \dots n$
\begin{equation}
\E^{-it M_{01}} \left( \E^{is M_{0j}} A \E^{-is M_{0j}} \right) \, \Omega = 
\E^{is(\ch(t) M_{0j} - \sh(t) M_{1j} )} \, \E^{-it M_{01}} A \,
\Omega. \label{6.3}
\end{equation}
According to the geodesic KMS--property of the vacuum and 
modular theory \cite{KaRi}, the vector--valued functions 
\begin{equation}
t \rightarrow \E^{-it M_{01}} B \, \Omega, \quad B \in {\CA}(\CW_1), 
\end{equation}
can be analytically continued into the strip 
$\{ z \in \CC : 0 > \mbox{Im} z > - \beta/2 \}$ and have continuous boundary
values at $z = -i\beta/2$, given by 
\begin{equation}
\E^{- (\beta/2) M_{01}} B \, \Omega = J_{\CW_1} B^{*} \, \Omega. \label{6.5}
\end{equation}
Moreover, for given $\gamma > \beta/2$ and sufficiently small $s$, the  
vector--valued function 
\begin{equation}
t \rightarrow \E^{-is(\ch(t) M_{0j} - \sh(t) M_{1j} )} \, \Phi, \quad |t|
< \gamma, 
\end{equation}
where $\Phi$ is any element of the dense set of analytic vectors 
described in Lemma \ref{2.3}, can 
be analytically continued into 
the complex circle $\{ z \in \CC : |z| < \gamma \}$. The 
continuation is given by 
\begin{equation}
\E^{-is(\ch(z) M_{0j} - \sh(z) M_{1j} )} \, \Phi,
\end{equation}  
where the exponential function is defined in the sense of power
series. Taking scalar products of the vectors in equation
\ref{6.3} with $\Phi$ we therefore obtain for sufficiently small $s$ 
by analytic continuation in $t$ the equality 
\begin{equation} 
(\Phi, \E^{- iz M_{01}}  \E^{is M_{0j}} A  \, \Omega) = 
(\E^{-is(\ch(\overline{z} ) M_{0j} - \sh(\overline{z} ) M_{1j} )} \,\Phi, 
\E^{- iz M_{01}}  A \, \Omega) \label{6.8a}
\end{equation} 
for 
$z$ in $\{ z \in \CC : |z| < \gamma \} \cap 
\{ z \in \CC : 0 > \mbox{Im} z > - \beta/2 \}$. 
Proceeding to the boundary point $z = - i\beta/2$ and making use of  
relations (\ref{6.2}) and (\ref{6.5}) we arrive at
\begin{equation} 
(\Phi, J_{W_1} \E^{is M_{0j}} A^*  \, \Omega) = 
(\E^{-is(\ch(i\beta/2) M_{0j} - \sh(i\beta/2) M_{1j} )} \,\Phi, 
J_{W_1} A^* \, \Omega). \label{6.8}
\end{equation} 
Since the operators $J_{W_1}$ and $\E^{is M_{0j}}$ are (anti--)unitary and 
the vectors $\Phi$ and $A^* \, \Omega$ are arbitrary elements of 
two dense sets in $\CH$ we conclude that  
$\E^{-is(\ch(i\beta/2) M_{0j} - \sh(i\beta/2) M_{1j} )}$ has to be 
unitary. If $s \neq 0$ this is only possible if $\beta$ is an integer 
multiple of $2 \pi$. As a matter of fact there holds $\beta = 2 \pi$ as we will
show next. 

If $\beta \geq 4 \pi$ then $2 \pi \leq \beta/2$ and hence 
the vectors in ${\CA}(\CW_1) \, \Omega$ are in the
domain of $\E^{- 2 \pi M_{01}}$. Now from (\ref{6.8a}) we see that   
for $A \in \CA(\CO)$ and sufficiently small $s,t$ such that  
$\E^{is M_{0j}} \E^{it M_{01}} A \E^{-it M_{01}} \E^{-is M_{0j}} \in
{\CA}(\CW_1)$ there holds 
\begin{equation} 
 \E^{- \pi M_{01}}  \E^{is M_{0j}} \E^{it M_{01}} A  \, \Omega = 
\E^{-is M_{0j}}  \E^{it M_{01}} \E^{- \pi M_{01}}  A \, \Omega.  
\end{equation}
By multiplication of this equation 
with the spectral projections $ P(\Delta)$ of $M_{01}$, where
$\Delta \subset \RR$ is compact, we proceed to  
\begin{equation} 
\E^{- \pi M_{01}} P(\Delta)\, \E^{is M_{0j}} \E^{it M_{01}} A  \, \Omega = 
 P(\Delta) \, \E^{-is M_{0j}}  \E^{it M_{01}} \E^{- \pi M_{01}}  A \, \Omega.  
\end{equation}
Since $\E^{- \pi M_{01}} P(\Delta)$ is a bounded operator 
the vector--valued functions on both sides of this equality can be 
analytically continued in $t$
into the strip $\{ z \in \CC : 0 < \mbox{Im} z < \pi \}$.  Therefore
the equality holds for all $t \in \RR$ and consequently 
\begin{equation} 
 \E^{- \pi M_{01}} P(\Delta) \, \E^{is M_{0j}}  P(\Delta) \, A  \, \Omega = 
 P(\Delta) \, \E^{-is M_{0j}}  P(\Delta) \, \E^{- \pi M_{01}}  A \, \Omega.  
\end{equation} 
As $\CA (\CO) \, \Omega$ is dense in $\CH$ we get 
$ \E^{- \pi M_{01}} P(\Delta) \, \E^{is M_{0j}}  P(\Delta) = 
P(\Delta) \, \E^{-is M_{0j}}  P(\Delta) \, \E^{- \pi M_{01}}$ 
on all spectral subspaces of $M_{01}$. So by left multiplication
of this equation with $\E^{- \pi M_{01}}$ we conclude that 
$P(\Delta) \, \E^{is M_{0j}}  P(\Delta)$ and $\E^{- 2 \pi M_{01}}$
commute. Since $\Delta$ was arbitrary, it follows that 
$ \E^{it M_{01}}$ and $\E^{is M_{0j}}$ commute for $j=2, \dots n$
which is only possible if $U$ is the trivial 
representation. So we have proved:
 
\begin{Theorem}
The geodesic temperature has 
the Gibbons--Hawking value, $\beta = 2 \pi $.
\end{Theorem}
With the help of relation (\ref{6.8}) we will now compute the adjoint
action of the modular conjugation
$J_{\CW_1}$ on $U(\dSG)$. As $\beta = 2 \pi$ we obtain  
from (\ref{6.8}) for small $s$  
\begin{equation} 
J_{W_1} \E^{is M_{0j}} = \E^{-is M_{0j}} J_{W_1}, \quad j = 2, \dots n, 
\end{equation} 
and it is then apparent that this relation holds for arbitrary $s \in
\RR$. The modular theory, on the other hand, implies that   
\begin{equation} 
J_{W_1} \E^{is M_{01}} = \E^{is M_{01}} J_{W_1}.
\end{equation} 
Since the boost operators $\E^{is M_{0j}}, \, j = 1, \dots n$, 
generate $U(\dSG)$, the adjoint action of $J_{W_1}$  on 
this group can 
be read off from these relations. After a moments reflection one sees 
that $J_{W_1}$ 
is an anti--unitary representer of the element $T \! P_1 \in O(1,n)$, 
where $T$ denotes time reflection and $P_1$ the reflection along the
spatial $1$-direction in the chosen coordinate system. Moreover, from 
Proposition \ref{6.1}, applied to $J_{W_1}$, 
and the preceding two equalities one obtains 
\begin{equation} 
J_{W_1} \AW J_{W_1} = \CA ( T \! P_1 \CW) \ \ \mbox{for} \ \ \CW
\subset \dS. 
\end{equation} 
Summarizing these results, we have established the following version 
of a PCT--Theorem in de Sitter space. 

\begin{Theorem} The modular conjugation $J_{W_1}$ associated with the 
wedge $\CW_1$ is an anti--unitary representer of 
the reflection $T \! P_1 \in O(1,n)$
which induces the corresponding action on $U(\dSG)$ and on 
the wedge algebras.
\end{Theorem}
An analogous result for wedges other than $\CW_1$ is  
obtained by de Sitter covariance. 
  
\section{Conclusions}
\setcounter{equation}{0}
Starting from the physically meaningful assumption that vacuum states 
in de Sitter space look like
equilibrium states for all geodesic observers with an {\em a priori} 
arbitrary temperature, we have analysed in a general setting 
the global structure of these
states. It turned out that they have essentially the same
properties as vacuum states in Minkowski space except that they are
not ground states.   

For mixed vacuum states it follows from 
the results of Sec.\ 4 that the respective 
sub-ensembles belong to different superselection sectors 
(phases) which can be
distinguished by elements of the center of the algebra of 
observables.  
By central decomposition one can always proceed to pure vacuum
states which are weakly mixing. 
It of interest in this context 
that this central decomposition can be performed by any
geodesic observer. As has been shown in Sec.\ 5,   
the relevant macroscopic order parameters can be constructed in every wedge
by suitable ``time averages'' of local observables. 

The geodesic temperature has 
the value predicted by Gibbons and Hawking also in the present 
general setting. 
This result could be  established without any further 
``stability assumptions'' by making use of 
the analytic structure of the de Sitter group, which was also 
essential in the proof of an analogue of the PCT--Theorem in de 
Sitter space. 

Our results provide evidence to the effect that 
the vacuum states, as defined in the present investigation,
indeed describe the envisaged physical situation. It would therefore 
be of interest to clarify the relation between our setting and 
the apparently more restrictive  
framework of maximal analyticity, proposed in \cite{BrEpMo} to
characterize vacuum states in de Sitter space.
\\[5mm] 

{\noindent \bf \Large Acknowledgements}\\[2mm]
We are grateful to J.\ Bros, U.\ Moschella, S.J.\ Summers and R.\ Verch for 
stimulating discussions on this issue and to the Erwin Schr\"odinger 
Institut in Vienna for hospitality and financial support provided 
during the program ``Local Quantum Physics'' in the fall of 1997.

\end{document}